\documentclass[runningheads]{llncs}
\usepackage[T1]{fontenc}
\usepackage{graphicx}
\usepackage[colorlinks=true, urlcolor=blue, linkcolor=blue, citecolor=blue]{hyperref}
\usepackage{xcolor}

\usepackage[caption=false]{subfig}

\usepackage{tikz}
\usetikzlibrary{positioning}
\usepackage{svg}

\usepackage{amsmath}

\usepackage{xurl}

\usetikzlibrary{arrows.meta, calc, decorations.pathreplacing}

\begin{document}
\title{Secure Wi-Fi Ranging Today: Security and Adoption of IEEE 802.11az/bk}
%
%
%
\author{Nikola Antonijevi\'c\inst{1} \and 
Bernhard Etzlinger\inst{2} \and 
Dave Singel\'ee\inst{3} \and 
Bart Preneel\inst{1}} 

\authorrunning{N. Antonijevi\'c et al.}

\institute{
COSIC, KU Leuven, Belgium\\
\email{\{nikola.antonijevic,bart.preneel\}@esat.kuleuven.be}
\and
Institute for Communications Engineering and RF-Systems, Johannes Kepler University Linz, Austria\\
\email{bernhard.etzlinger@jku.at}
\and
DistriNet (Group T), KU Leuven, Belgium\\
\email{dave.singelee@kuleuven.be}
}

\maketitle              
\begin{abstract}

Ranging and localisation have become critical for many applications and services. The Wi-Fi (IEEE~802.11) standard is a natural candidate for providing these functions across diverse environments, given its widespread deployment. The IEEE~802.11az amendment, finalised in 2023, introduces ``Next Generation Positioning'' mechanisms to secure and harden the existing insecure Wi-Fi Fine Timing Measurement (FTM) ranging solution. Moreover, the recent IEEE~802.11bk amendment increases the available bandwidth with the goal of approaching the centimetre-level ranging accuracy of ultra-wideband (UWB) systems. This paper examines to what extent these promises hold from a security and deployability perspective.
 
We analyse the core mechanisms of secure Wi-Fi ranging as defined in IEEE~802.11az and IEEE~802.11bk at both the logical and physical layers, combining standards analysis with simulations and measurements on commercial and development hardware. At the logical layer, we show how common deployment choices can result in unauthenticated ranging, downgrade attacks, and simple denial-of-service attacks, making it difficult to securely realise many high-stakes use cases. At the physical layer, we study the predictability of secure ranging waveforms, the security impact of symbol repetition, and how waveform design choices affect compliance with spectral masks under realistic RF behaviour.
 
Our results show that secure Wi-Fi ranging is highly sensitive to configuration choices and is non-trivial to implement on existing hardware. This is also evidenced by the currently limited support for secure \mbox{Wi-Fi} ranging in commodity devices. This paper provides practical guidelines for using secure FTM safely and recommendations to vendors and standardisation bodies to improve its robustness and deployability.

\end{abstract}
\section{Introduction}

The family of wireless network protocols based on IEEE~802.11, commonly known as \mbox{Wi-Fi}, underpins connectivity for billions of devices worldwide. For an increasing number of applications, however, a device's location is as important as its ability to communicate: precise knowledge of \emph{where} a device is can determine which services it can access and how it can be used. Since satellite-based systems such as GPS perform poorly indoors, \mbox{Wi-Fi} has emerged as an attractive platform for indoor ranging and localisation, leveraging infrastructure that is already widely deployed~\cite{wifi_positioning_survey}. Typical examples include navigation in large buildings, asset and inventory tracking, access control and automation in smart homes and offices, and applications that depend on trustworthy proximity information~\cite{standard_80211_2024}.

The \mbox{Wi-Fi} Fine Timing Measurement (FTM) was first introduced in the IEEE~802.11mc amendment in 2016, enabling more precise \mbox{Wi-Fi} ranging by using Time of Flight (ToF) measurements to estimate the distance between two stations~\cite{IEEE80211az}. Prior \mbox{Wi-Fi} positioning systems mostly relied on received signal strength (RSS), with no standardised ToF-based ranging. FTM significantly improved the accuracy and usability of \mbox{Wi-Fi}-based ranging, but the original procedures did not protect the integrity or authenticity of ranging messages. Subsequent research by Schepers et al.~\cite{wifi_80211mc_security1} showed that legacy FTM is highly vulnerable to distance manipulation and spoofing attacks. To address these weaknesses, the IEEE~802.11az amendment was released in 2023. This \mbox{Wi-Fi} Next Generation Positioning (NGP) standard strengthens FTM at both the logical and physical layers, for example, through protected management frames and secure high-efficiency long training fields (HE-LTFs). Most recently, IEEE~802.11bk further extended FTM by increasing the maximum bandwidth to 320~MHz to approach the ranging accuracy of ultra-wideband (UWB), while inheriting the security mechanisms of IEEE~802.11az~\cite{standard_80211bk}. The resulting procedures are now consolidated in the IEEE~802.11-2024 standard~\cite{standard_80211_2024}.

Despite these developments, commercial support for secure FTM as defined in IEEE~802.11az remains limited. Only a small number of devices and development platforms publicly advertise secure FTM capabilities, and even fewer expose them in a documented, developer-accessible way. As a result, the security and deployability of secure FTM in IEEE~802.11az, and the implications for IEEE~802.11bk, have not been examined in depth. Existing security analyses focus on legacy FTM in IEEE~802.11mc and largely treat the physical layer as an idealised ranging primitive, without considering the concrete secure HE-LTF waveform or the availability of secure FTM features in deployed hardware.

This paper aims to fill this gap. We analyse the most important logical and physical-layer mechanisms of secure FTM in IEEE~802.11az and their relevance for IEEE~802.11bk. Our analysis combines a close reading of the IEEE~802.11-2024 standard with simulations in a MATLAB-based IEEE~802.11az framework and measurements on development boards that claim to support secure FTM. To the best of our knowledge, this is the first security-focused assessment of \mbox{Wi-Fi} Next Generation Positioning that jointly considers logical and physical mechanisms and their practical realisation.

\subsubsection{Contributions.} Our main contributions can be summarised as follows:

\begin{itemize}
  \item \textbf{Issue-driven analysis of secure FTM.}
  We analyse core logical and physical-layer mechanisms of secure FTM in IEEE~802.11az and identify several classes of issues, including fragile trust assumptions in common WPA2/WPA3-Personal and PASN deployments, downgrade and denial-of-service risks, and failure modes related to secure HE-LTF counters and symbol repetition.

  \item \textbf{Security assessment of the secure FTM waveform.}
  Using a MATLAB-based model calibrated with measurements from a commercial platform, we study the predictability of secure HE-LTF waveforms under partial observation and quantify how much an attacker can bias time of arrival estimates, and thus the measured range, with advanced replicas without breaking demodulation or violating spectral constraints.

  \item \textbf{Deployment insights and recommendations.}
  We combine a snapshot of current IEEE~802.11az support in commercial and development hardware with our technical findings to explain why secure \mbox{Wi-Fi} ranging remains sparsely deployed. Based on this, we provide recommendations for device vendors and standardisation bodies on configuration choices and physical-layer constraints that future profiles and amendments should address.
\end{itemize}

The rest of this paper is organised as follows. Section~\ref{sec:background} provides background on legacy and secure FTM and surveys current support. Section~\ref{sec:methodology-scope} describes our threat model and methodology. Section~\ref{sec:logical} and Sect.~\ref{sec:physical} present our logical-layer and physical-layer analyses. Section~\ref{sec:discussion} discusses the implications of our findings and recommendations, and Sect.~\ref{sec:related-work} positions our work in the existing literature before we conclude the paper in Sect.~\ref{sec:conclusion}.

\section{Background and Support for Secure FTM}\label{sec:background}

This section introduces legacy \mbox{Wi-Fi} Fine Timing Measurement (FTM), also known as \mbox{Wi-Fi} Round Trip Time (RTT), and then summarises the secure FTM extensions in IEEE~802.11az/bk and their current adoption~\cite{standard_80211_2024,standard_80211bk}.

\subsection{Fine Timing Measurement (802.11mc)}\label{sec:11mc}

FTM, introduced in IEEE~802.11mc in 2016, enables \mbox{Wi-Fi} ranging by using Time of Flight (ToF) measurements instead of received signal strength (RSS), significantly improving indoor distance estimation accuracy~\cite{wifi_positioning_survey}. Two parties, an \emph{initiator} and a \emph{responder}, engage in an FTM session that consists of negotiation, measurement exchange, and termination. Figure~\ref{fig:ftm-session} illustrates a typical exchange between an initiator station (STA) and a responder access point (AP).

\vspace{-2mm}
\subsubsection{Negotiation.}
The initiator station (ISTA) discovers an FTM-capable responding station (RSTA) from a capability bit in the AP beacon and starts negotiation by sending an FTM Request frame (IFTMR). This request selects the measurement procedure and configures RF/PHY and protocol parameters (e.g., channel bandwidth and burst settings). Once the responder acknowledges this frame, the ISTA is ready to enter the measurement exchange phase.

\vspace{-2mm}
\subsubsection{Measurement exchange.}
\begin{figure}[t]
  \centering
  \scalebox{0.51}{
\begin{tikzpicture}[
    >=Latex, 
    font=\sffamily, 
    thick, 
    entity/.style={draw, line width=1.5pt, minimum width=2cm, minimum height=1cm, font=\Large\bfseries},
    msg/.style={->, line width=1.5pt},
    dashed_msg/.style={->, line width=1.5pt, dashed}
]

\def\apX{0}
\def\staX{7}
\def\timeStart{0}
\def\timeEnd{-13} 

\node[entity] (ap) at (\apX, \timeStart) {AP};
\node[entity] (sta) at (\staX, \timeStart) {STA};

\draw[line width=1.5pt] (ap.south) -- (\apX, \timeEnd);
\draw[line width=1.5pt] (sta.south) -- (\staX, \timeEnd);

\node[anchor=south, font=\large] at (-2.5, -3.5) {Time};
\draw[->, line width=2.5pt] (-2.5, -3.5) -- (-2.5, -9.5);

\filldraw[fill=white, line width=1.5pt] 
    (-0.5, -1.65) -- (7.5, -1.65)   
    -- (7.5, -1.35)                 
    -- (8.5, -2.00)                 
    -- (7.5, -2.65)                 
    -- (7.5, -2.35)                 
    -- (-0.5, -2.35)                
    -- (-0.5, -2.65)                
    -- (-1.5, -2.00)                
    -- (-0.5, -1.35)                
    -- cycle;

\node[font=\large] at (3.5, -2.0) {STA detects FTM capable AP};

\def\yReqStart{-4.0}
\def\yReqEnd{-4.5}
\draw[dashed_msg] (\staX, \yReqStart) -- (\apX, \yReqEnd) node[midway, above, sloped] {FTM Request};

\def\yAckStart{-5.0}
\def\yAckEnd{-5.5}
\draw[dashed_msg] (\apX, \yAckStart) -- (\staX, \yAckEnd) node[midway, above, sloped] {ACK};

\def\yFtmStart{-7.0}
\def\yFtmEnd{-8.0}
\draw[msg] (\apX, \yFtmStart) coordinate (t1_pos) -- (\staX, \yFtmEnd) coordinate (t2_pos) node[midway, above, sloped] {FTM};

\node[left, font=\Large] at (t1_pos) {$t_1$};
\node[right, font=\Large] at (t2_pos) {$t_2$};

\def\yAckMStart{-9.0}
\def\yAckMEnd{-10.0}
\draw[msg] (\staX, \yAckMStart) coordinate (t3_pos) -- (\apX, \yAckMEnd) coordinate (t4_pos) node[midway, above, sloped] {ACK};

\node[right, font=\Large] at (t3_pos) {$t_3$};
\node[left, font=\Large] at (t4_pos) {$t_4$};

\draw[decorate, decoration={brace, amplitude=12pt}, line width=1.5pt]
    (\staX+1.2, \yFtmEnd+0.5) -- (\staX+1.2, \yAckMStart-0.5)
    node[midway, right=20pt, align=left, font=\large] {Repeated\\[0.5em] N times};

\def\yFinalStart{-11.5}
\def\yFinalEnd{-12.2}
\draw[msg] (\apX, \yFinalStart) -- (\staX, \yFinalEnd) node[midway, above, sloped, font=\Large] {$t_1, t_4$};


\end{tikzpicture}
}
  \caption{Example of a typical FTM ranging session between the station (STA) and the access point (AP)\@.}
  \label{fig:ftm-session}
  \vspace{-4mm}
\end{figure}
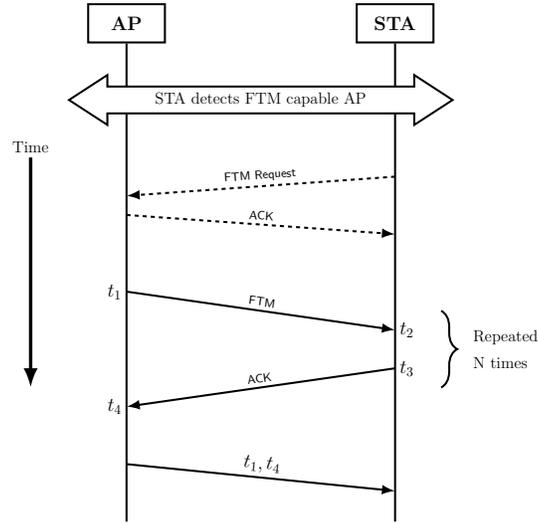

During measurement exchange, the AP and STA send FTM message pairs with embedded transmit and receive timestamps. The AP transmits an FTM frame at time $t_1$ (Fig.~\ref{fig:ftm-session}), containing standard IEEE~802.11 training fields that serve both demodulation and Time of Arrival (ToA) estimation. The initiator records the receive time $t_2$, responds with an ACK at time $t_3$, and the AP receives it at time $t_4$. The AP later reports $t_1$ and $t_4$ in Location Measurement Report (LMR) frames.

The transmit timestamps $t_1$ and $t_3$ are scheduled transmission instants, whereas $t_2$ and $t_4$ require ToA estimation, that is, identifying the earliest detectable signal component. Various high-resolution ToA estimators have been proposed, including MUSIC-based approaches~\cite{toa_estimation_music}. Using the four timestamps, the initiator computes the round-trip time and corresponding range $d = \frac{c}{2}\bigl[(t_4 - t_1) - (t_3 - t_2)\bigr]$, assuming propagation at the speed of light $c$. In practice, this is repeated over multiple FTM bursts and the distance estimates are averaged.

\vspace{-2mm}
\subsubsection{Termination.}
An FTM session terminates after the final burst instance, or earlier if the ISTA or RSTA transmits control values that signal termination.

\subsection{Secure Fine Timing Measurement (802.11az and 802.11bk)}\label{sec:bg_secureftm}

In 2023, the IEEE~802.11az amendment was released with the primary goal of hardening existing FTM procedures. Legacy FTM offers little integrity or authenticity protection: timestamps are sent in the clear, management frames are unencrypted, and ToA estimation training fields follow a public, deterministic structure. In multicarrier orthogonal frequency-division multiplexing (OFDM) systems like Wi-Fi, this predictability enables powerful ranging and location attacks~\cite{multicarrier_security}. NGP addresses these weaknesses by adding MAC and physical-layer protections. IEEE~802.11bk largely reuses these mechanisms while extending the available bandwidth and improving ranging accuracy.

\subsubsection{Protected Management Frames.}
Protected Management Frames (PMF), originally introduced in IEEE~802.11w and later integrated into the main standard~\cite{standard_80211_2024}, ensure that selected management frames are cryptographically protected after key establishment. NGP adds FTM-related management frames to this protection set. In a typical FTM session (Fig.~\ref{fig:ftm-session}), the FTM Request and the frames carrying $t_1$ and $t_4$ are protected at the MAC layer. This provides authenticity and confidentiality, preventing an active attacker from modifying timestamps or injecting forged messages without detection.

\subsubsection{Preassociation Security Negotiation.}\label{sec:pasn}

Before IEEE~802.11az, stations authenticated and derived shared keys through mechanisms such as WPA2-Personal, 802.1X-based Enterprise authentication, and WPA3 schemes including Simultaneous Authentication of Equals (SAE). These mechanisms are coupled to association and are unsuited for lightweight ranging without a full data connection.

To decouple secure ranging from full association, IEEE~802.11az introduces preassociation security negotiation (PASN). PASN derives a fresh shared key for preassociation exchanges using authentication frames so that selected management traffic, including FTM frames, can be protected even when no data-encryption keys exist.

PASN can operate in two modes~\cite{standard_80211_2024}. When run on top of an existing authenticated \mbox{Wi-Fi} connection (for example a WPA3 network with established data keys), it reuses that security context to derive a separate key for preassociation exchanges, inheriting mutual authentication from the underlying protocol. When used on its own, without a prior authenticated association, PASN can still derive keys and protect management frames, but the standard explicitly does not guarantee mutual authentication. This lightweight mode reduces overhead for ranging, but it has important security implications that we analyse in Sect.~\ref{sec:pasn_sec}.

\subsubsection{Secure LTF sequence.}
The IEEE~802.11ax PHY defines a high efficiency (HE) null data packet (NDP), a physical layer protocol data unit (PPDU) that carries no payload and consists only of preamble and training fields. In ranging mode, this HE ranging NDP is used solely for synchronisation, signalling, channel estimation, and ToA estimation. Within this preamble, the high efficiency long training fields (HE-LTFs) form the main OFDM training sequence that the receiver uses for ToA estimation.

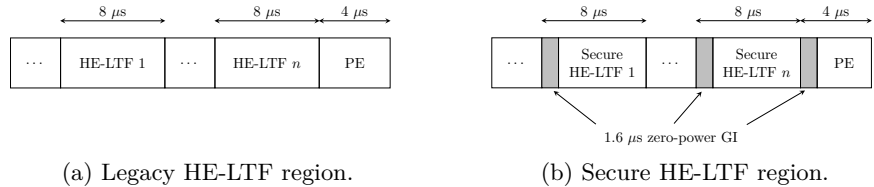
\begin{figure}[t]
  \centering
  \subfloat[Legacy HE-LTF region.\label{fig:LTF_mc}]{%
    \begin{minipage}[b]{0.48\linewidth}
      \centering
      \resizebox{0.9\linewidth}{!}{\scalebox{0.7}{
\begin{tikzpicture}[
    >={stealth},
    line width=0.6pt,
    field/.style={rectangle, draw=black, minimum height=1.2cm, inner sep=0pt, font=\normalsize},
    timing/.style={midway, font=\normalsize}
]


\node[field, minimum width=1.2cm] (hestf) at (0,0) {\dots};
\node[field, minimum width=2.5cm, right=-\pgflinewidth of hestf] (heltf1) {HE-LTF 1};
\node[field, minimum width=1.2cm, right=-\pgflinewidth of heltf1] (dots) {\dots};
\node[field, minimum width=2.5cm, right=-\pgflinewidth of dots] (heltfn) {HE-LTF $n$};
\node[field, minimum width=1.7cm, right=-\pgflinewidth of heltfn] (pe) {PE};


\def\arroffset{0.4cm} 


\draw[<->] ([yshift=\arroffset]heltf1.north west) -- ([yshift=\arroffset]heltf1.north east)
node[timing, anchor=base, yshift=3pt] {8 $\mu$s};


\draw[<->] ([yshift=\arroffset]heltfn.north west) -- ([yshift=\arroffset]heltfn.north east)node[timing, anchor=base, yshift=3pt] {8 $\mu$s};

\draw[<->] ([yshift=\arroffset]pe.north west) -- ([yshift=\arroffset]pe.north east) 
    node[timing, anchor=base, yshift=3pt] {4 $\mu$s};

\end{tikzpicture}
}}
      \vspace{0.81cm}
    \end{minipage}%
  }\hfill
  \subfloat[Secure HE-LTF region.\label{fig:LTF_az}]{%
    \begin{minipage}[b]{0.48\linewidth}
      \centering
      \resizebox{0.9\linewidth}{!}{\scalebox{0.7}{
\begin{tikzpicture}[
    >={stealth},
    line width=0.6pt,
    field/.style={rectangle, draw=black, minimum height=1.2cm, inner sep=0pt, font=\normalsize},
    gi/.style={rectangle, draw=black, fill=black!25, minimum height=1.2cm, inner sep=0pt},
    timing/.style={midway, font=\normalsize}
]


\node[field, minimum width=1.2cm] (hestf) at (0,0) {\dots};

\node[gi, minimum width=0.4cm, right=-\pgflinewidth of hestf] (gi1) {};
\node[field, minimum width=2.1cm, right=-\pgflinewidth of gi1, align=center] (heltf1) {Secure\\HE-LTF 1};

\node[field, minimum width=1.2cm, right=-\pgflinewidth of heltf1] (dots) {\dots};

\node[gi, minimum width=0.4cm, right=-\pgflinewidth of dots] (gin) {};
\node[field, minimum width=2.1cm, right=-\pgflinewidth of gin, align=center] (heltfn) {Secure\\HE-LTF $n$};

\node[gi, minimum width=0.4cm, right=-\pgflinewidth of heltfn] (gipe) {};
\node[field, minimum width=1.3cm, right=-\pgflinewidth of gipe] (pe) {PE};

\def\arroffset{0.4cm} 

\draw[<->] ([yshift=\arroffset]gi1.north west) -- ([yshift=\arroffset]heltf1.north east)
node[timing, anchor=base, yshift=3pt] {8 $\mu$s};

\draw[<->] ([yshift=\arroffset]gin.north west) -- ([yshift=\arroffset]heltfn.north east)
node[timing, anchor=base, yshift=3pt] {8 $\mu$s};

\draw[<->] ([yshift=\arroffset]gipe.north west) -- ([yshift=\arroffset]pe.north east) 
    node[timing, anchor=base, yshift=3pt] {4 $\mu$s};

\node[font=\normalsize] (gibot) at ([yshift=-1.2cm]dots.south) {1.6 $\mu$s zero-power GI};


\draw[->, shorten >=3pt, shorten <=3pt] (gibot.north west) -- (gi1.south);
\draw[->, shorten >=3pt, shorten <=3pt] (gibot.north) -- (gin.south);
\draw[->, shorten >=3pt, shorten <=3pt] (gibot.north east) -- (gipe.south);

\end{tikzpicture}
}}
    \end{minipage}%
  }
  \caption{Comparison of the HE-LTF part of an HE ranging NDP in the legacy (non-secure) and secure IEEE~802.11az formats. Earlier preamble fields and headers are omitted for clarity.}
  \vspace{-4mm}
  \label{fig:LTF_format}
\end{figure}

Figure~\ref{fig:LTF_format} focuses on this HE-LTF portion of the ranging NDP. In the legacy, non-secure format (Fig.~\ref{fig:LTF_mc}), a series of HE-LTF symbols with publicly specified deterministic patterns is transmitted back to back. This simplifies receiver design but makes channel estimates and ToA measurements susceptible to manipulation and signal-processing-based attacks~\cite{multicarrier_security}.


IEEE~802.11az replaces these fixed patterns with \textit{secure HE-LTFs} (Fig.~\ref{fig:LTF_az}), where training content is derived from secret keying material. Pairwise key material shared between initiator and responder is used to generate a pseudorandom bit stream that determines the HE-LTF symbols. Concretely, a key derivation key (KDK) embedded in the pairwise temporal key (PTK), established via PASN or conventional WPA2/WPA3 authentication, is first used to derive a long-term \emph{secure LTF key seed} for the lifetime of the PTK. Each secure FTM measurement instance is associated with a \emph{secure LTF counter}, initialised when the first secure session starts and incremented for every subsequent instance. A key derivation function combines the secure LTF key seed and the counter to produce per-measurement material: a Validation Sequence Authentication Code (SAC) and the keys used to generate the secure HE-LTF waveforms at the initiator and responder. The resulting bits are mapped onto 64-QAM symbols on the active subcarriers, yielding a pseudorandom training waveform. The secure LTF counter is intended to ensure that each secure HE-LTF instance is used at most once, while the Validation SAC allows the receiver to verify that the observed training waveform corresponds to the expected counter value and key.

In addition, IEEE~802.11az changes the guard interval between successive HE-LTF symbols. In the legacy format, this interval is implemented as a cyclic prefix, that is, a copy of the end of the OFDM symbol. In the secure HE-LTF format, the guard interval is forced to zero power. Removing the deterministic repetition of the cyclic prefix and using a zero-power guard interval reduces redundancy in the waveform and limits opportunities for an attacker to exploit structural correlation in FTM. We return to the implementation implications of this design choice in Sect.~\ref{sec:zero-power-gi}.

\subsection{Adoption of Secure FTM}\label{sec:support}

Since the release of IEEE~802.11mc in 2016, a broad range of products have added support for FTM. Google has been a major driver of this adoption: Wi-Fi RTT support is exposed to applications through the Android Wi-Fi RTT API~\cite{wifi_rtt}, and tools such as WiFiRttScan provide convenient testing~\cite{wifirttlocator}. Most Google Pixel phones and several other Android devices support \mbox{Wi-Fi} RTT~\cite{android_wifi_rtt}. A variety of APs and development boards implement the protocol, and FTM has seen practical deployment in indoor location services, for example by IndoorAtlas~\cite{indooratlas}.

In contrast, support for secure FTM as specified in IEEE~802.11az remains limited. Although the Android Wi-Fi RTT stack documents secure FTM modes, devices that actually implement these features in hardware and firmware appear to be scarce~\cite{wifi_rtt}. Even recent Google hardware, such as Nest Wifi Pro access points and Pixel smartphones, does not yet expose IEEE~802.11az capabilities in practice~\cite{android_wifi_rtt}, and our own tests with a Google Pixel~9a and the WiFiRttLocator app did not reveal any secure FTM functionality. Public information currently mentions only a small number of enterprise-grade access points, such as the Aruba AP-734, as advertising support for secure FTM~\cite{android_wifi_rtt}.

Several vendors also advertise development boards that claim IEEE~802.11az support and promise fine-grained control over PHY and MAC parameters. For this work, we relied on one such platform, which we cannot name due to a non-disclosure agreement (NDA). Despite the vendor's claim of full IEEE~802.11az compliance, we found that only a subset of secure FTM features was available in practice; the board implemented protected management frames but not secure HE-LTFs or other physical layer protections. We return to these limitations in Sect.~\ref{sec:scope}, where we explain how they influenced our experimental design.

\section{Threat Model and Methodology}
\label{sec:methodology-scope}

Our analysis of secure FTM combines a close reading of the IEEE~802.11-2024 standard with simulations and hardware measurements to evaluate the protocol against a powerful proximal adversary. 

\subsection{Threat Model}\label{sec:threat-model}
In this work, we assume that the two parties who wish to determine their mutual distance are honest. Otherwise, falsifying measurements is trivial since endpoints report their own timestamps (Sect.~\ref{sec:11mc}). The adversary is a proximal third party aiming to manipulate the measured distance to bypass access controls or location restrictions. We assume a powerful active adversary in the Dolev-Yao style~\cite{dolev2003security} who can arbitrarily observe and manipulate wireless traffic using high-power radios and arbitrary baseband waveforms. However, the attacker cannot compromise the honest stations' internal state, break standard cryptographic primitives, or violate physical limits like the speed of light. They possess complete knowledge of the IEEE~802.11-2024 standard and its secure FTM amendments~\cite{standard_80211_2024}.

\subsection{Scope and limitations}\label{sec:scope}
We focus on the secure FTM variants in IEEE~802.11az and their implications for IEEE~802.11bk, which retains the same security mechanisms over wider bandwidths. Rather than a formal proof, we perform an issue-driven analysis highlighting concrete weaknesses, limitations, and implementation pitfalls in realistic deployments using simulations and experiments.

\subsubsection{Simulation.}
We use MATLAB's IEEE~802.11az waveform and positioning examples~\cite{mathworks80211azGen,mathworks80211azToA}, which implement draft IEEE~P802.11az/D2.0, to parameterise and generate ranging NDP waveforms and to estimate ToA using a MUSIC super-resolution approach. We extend this baseline framework to analyse both secure and non-secure waveforms and to study ToA estimation under various channel and attacker strategies.

\subsubsection{Experiments.}
Testing secure FTM in practice is challenging due to the limited availability of NGP hardware (Sect.~\ref{sec:support}). We relied on a development board that explicitly advertised IEEE~802.11az support; however, it only implemented protected management frames. The manufacturer (redacted via NDA) confirmed that secure HE-LTFs and physical-layer protections are not available and not planned for the current chipset due to architectural constraints (discussed further in Sect.~\ref{sec:dis_support}). Despite this, we used the boards to perform FTM-based ranging, recorded the transmitted waveforms, and compared them against our simulations to establish a non-secure FTM baseline. We also cross-checked IEEE~802.11mc ranging against a commercial Google Pixel~9a. These traces and simulations form the core dataset for our physical-layer analysis (Sect.~\ref{sec:physical}).

\section{Logical-Layer Analysis}\label{sec:logical}
In this section, we analyse the logical-layer mechanisms of secure FTM in NGP and how authentication, mode selection, and configuration affect the intended security goals.
We identify three main classes of issues.
First, common deployment modes such as WPA2/WPA3-Personal and standalone PASN fail to bind secure FTM to a uniquely trusted infrastructure entity, limiting the security of many proposed use cases (Sect.~\ref{sec:personal-mode} and Sect.~\ref{sec:pasn_sec}).
Second, several configuration and negotiation points admit downgrades from secure to legacy FTM or from secure HE-LTFs to deterministic training sequences, creating silent gaps between intended and effective security (Sect.~\ref{sec:critical-config}).
Third, when downgrades are prevented through strict policies, secure FTM becomes fragile from an availability perspective and is easily disrupted by simple denial-of-service attacks (Sect.~\ref{sec:dos}).

\subsection{Personal Mode Limitations}\label{sec:personal-mode}
NGP is positioned as an enabler for applications in public, home, and enterprise environments~\cite{IEEE80211az}.
In practice, many such deployments are expected to rely on WPA2-Personal or WPA3-Personal, or on other settings without a dedicated enterprise authentication infrastructure.
In this section, we examine how these Personal mode configurations affect the security of secure FTM and to what extent they can support security-critical use cases.

\subsubsection{Shared-passphrase risks in WPA2-Personal.}
WPA2-Personal security is based on a single pre-shared passphrase: all stations deterministically derive the same pairwise master key (PMK) from this passphrase and the network SSID.
During the four-way handshake, the PMK and fresh nonces are used to derive the PTK, which protects data, management frames, and is also used for generating the secure HE-LTFs employed for ranging~\cite{standard_80211_2024}.
Any party that knows the passphrase and observes a four-way handshake can compute the PTK for that association.
In many WPA2-Personal deployments this capability is realistic: home, retail, and small office networks commonly share the same passphrase with multiple users, guests, or devices, any of whom may act as an adversary.
Such an attacker can passively record the four-way handshake of a victim device and derive the PTK for that ongoing association as soon as the handshake completes.
With this PTK, the attacker can forge or modify protected management frames (for example, LMR) and generate valid secure HE-LTFs, thereby influencing ranging at the physical layer, as discussed in Sect.~\ref{sec:distance_attack}.
In effect, `secure' ranging in WPA2-Personal reduces to the secrecy of a widely shared passphrase, which is a weak foundation for security-critical distance decisions.

\subsubsection{Susceptibility of WPA3-Personal to active attacks.}
Compared to WPA2-Personal, WPA3-Personal strengthens the handshake by replacing the PSK-based four-way handshake with the Simultaneous Authentication of Equals (SAE) protocol, which provides forward secrecy~\cite{Dragonblood2019}.
An attacker who only knows the network password and passively records an SAE exchange between an honest access point and a client cannot reconstruct the resulting PTK, since the Diffie–Hellman components in SAE prevent recovery of the session key from the transcript and the password alone.
Thus, the passive attack described for WPA2-Personal does not directly apply to WPA3-Personal.

However, SAE only proves that both endpoints know the same password; it does not bind the handshake to a specific, trusted access point, a structural limitation shared with WPA2-Personal.
If an attacker knows the shared password, they can deploy a rogue access point (an ``evil twin'') and actively participate in the SAE exchange.
A victim device, unable to distinguish the rogue AP from the legitimate one at the cryptographic level, will complete the SAE handshake and derive a valid PTK with the attacker.
In 2024, work by Vanhoef and Robben~\cite{vanhoef2024_twin} shows that WPA3-PK, an extension to WPA3-Personal, adds public-key-based authentication for hotspots; they demonstrate in practice that any party who knows the password can create a rogue clone of the network and have clients accept it as the legitimate infrastructure.
Any secure FTM exchanges performed over this association will attest proximity to the attacker, not to the intended anchor point, defeating the purpose of proximity-based access control.
In this sense, WPA3-Personal provides stronger confidentiality against passive adversaries than WPA2-Personal, but still falls short of providing the strong binding that authenticated location services require when the password is shared widely.

\subsection{PASN}\label{sec:pasn_sec}

Managing shared credentials across heterogeneous devices in public or transient environments is operationally challenging, and requiring a full association for every ranging interaction is often impractical for guest navigation or \textit{ad hoc} use cases. To reduce this overhead, NGP introduces preassociation security negotiation (PASN), which allows two stations to derive keys and protect management and FTM frames, without establishing a full data connection~\cite{wifi_80211az_crowd}.

If PASN runs on top of an existing authenticated \mbox{Wi-Fi} connection, it reuses the established security context to derive fresh keys and essentially inherits the mutual authentication of the underlying WPA2 or WPA3 mechanism. The situation is different when PASN is used on its own, without a prior authenticated association: in this standalone setting, the standard explicitly does not guarantee mutual authentication between the two stations. PASN can still derive keys and protect management frames, but there is no strong binding between those keys and a long-term identifier.

As a consequence, standalone PASN is vulnerable to person-in-the-middle abuse. An active adversary can first perform a PASN exchange with the access point and the station separately, establishing independent keying material with each, and then remain on the path, relaying or altering protected management frames and secure HE-LTF content between them. By crafting and forwarding frames appropriately, the attacker can steer secure FTM ranging results to attest proximity either to itself or to the honest access point. Conceptually, this mirrors the Personal mode issues in Sect.~\ref{sec:personal-mode}: secure FTM is not firmly bound to a specific, trusted infrastructure entity, but only to whoever can participate in the key establishment. For high-stakes NGP use cases, this again points to the need for deployments rooted in robust mutual authentication, rather than relying solely on lightweight or password-based mechanisms.

\subsection{Critical configuration points}\label{sec:critical-config}
The security guarantees of NGP ultimately depend on how devices are configured. In practice, permissive or misaligned settings can significantly weaken the protection that secure FTM is meant to provide. Here, we highlight configuration choices that are particularly security-critical.

\subsubsection{Ranging mode selection and security gaps.}
As discussed in Sect.~\ref{sec:11mc}, an FTM procedure starts when an initiator (ISTA) transmits a Fine Timing Measurement Request (FTMR) frame to the responder (RSTA). This FTMR carries crucial configuration parameters, including the ranging mode.

IEEE~802.11az introduces several modes to improve scalability and efficiency: trigger-based (TB), non-trigger-based (non-TB), enhanced distributed channel access (EDCA)-based FTM exchange, and passive TB measurement~\cite{standard_80211_2024}. By default, stations use TB or non-TB ranging. The standard explicitly restricts EDCA-based ranging to special cases, such as 6~GHz band operation or when a station lacks TB/non-TB support, meaning EDCA acts as a legacy fallback.

This has direct security implications. Secure HE-LTFs with pseudorandom training content and zero-power guard intervals are defined only for TB and non-TB ranging. For EDCA-based ranging, secure HE-LTF mechanisms are not specified and are therefore not used. Even when MAC-level protection is in place, selecting EDCA-based ranging leaves the physical layer without the secure HE-LTF frames envisaged by NGP, creating a gap we revisit in Sect.~\ref{sec:physical}.

If devices are not configured carefully, an active attacker can exploit this as a mode-level downgrade. A proximal adversary can block an FTMR in which the ISTA requests TB or non-TB ranging and instead forge an FTMR towards the RSTA that selects EDCA-based ranging. The RSTA, following the standard rules and believing that TB or non-TB ranging is unavailable, replies with parameters for an EDCA-based session. From the ISTA's perspective, this appears as a legitimate negotiation where only the legacy compatible mode is supported, so it proceeds without secure HE-LTFs. In short, EDCA-based ranging remains a legacy path that can bypass secure FTM protections at the mode-selection stage.

\subsubsection{Negotiation-based downgrade from secure to legacy FTM.}
Beyond the choice of ranging mode, the FTMR also determines whether the session uses secure or legacy FTM. In IEEE~802.11az, this is reflected in whether the FTMR is sent as a protected management frame under an established PTK or as an unprotected management frame. If the ISTA has a PTK with the RSTA (a pairwise temporal key security association, or PTKSA) and sends a protected FTM Request Action frame as the initial FTMR, it signals intent to perform a secure FTM procedure. If the FTMR is sent unprotected, the ISTA initiates a non-secure, legacy FTM session.

The final mode is determined by the RSTA's policy. The standard allows an access point to require a PTKSA before accepting FTM, in which case it shall reject FTMR frames that are not protected under the PTK. In that configuration, only FTM sessions that run under a PTK are permitted. If the RSTA does not require a PTKSA, it may accept unprotected FTMR frames and proceed with a non-secure FTM procedure for compatibility with legacy devices.

Even when a PTKSA exists and the FTMR is protected, a second negotiation decides whether secure HE-LTFs are used. The ISTA indicates support, and optionally a requirement, for secure HE-LTFs by including the secure HE-LTF subelement in the Ranging Parameters element and setting the \emph{Secure HE-LTF Required} field. If the ISTA only signals support but does not set this field to 1, or if the RSTA does not echo the secure HE-LTF subelement with \emph{Secure HE-LTF Required} set to 1 in its response, the session proceeds with legacy, deterministic HE-LTFs even though control frames are protected at the MAC layer. From a security perspective, this is a downgrade of the ranging waveform to the non-secure baseline.

If these policies are permissive, the system is vulnerable to straightforward downgrades. An active attacker near the ISTA can block the first protected FTMR and then forge an equivalent unprotected FTMR towards the RSTA. An RSTA configured to accept both secure and non-secure FTM will respond with parameters for a non-secure session. From the ISTA's perspective, this is indistinguishable from a legitimate negotiation with a non-secure RSTA, so it proceeds with legacy FTM and predictable training sequences. Devices that support secure FTM but do not \emph{require} both a PTK and secure HE-LTFs therefore remain exposed to silent downgrades at either the MAC or physical layer. Conversely, configuring the RSTA to require a PTK for FTM and to honour the \emph{Secure HE-LTF Required} indication hardens the system, but reduces backward compatibility, since FTM is then only possible with peers that support protected management frames and secure HE-LTF procedures.

\subsubsection{Error recovery and counter reuse.}\label{sec:error-recovery}
The IEEE~802.11az standard specifies an explicit error recovery procedure for secure HE-LTF exchanges when a station detects an invalid parameter, for example when the received Validation SAC does not match the expected value. Conceptually, a failed measurement instance is marked invalid without advancing the Validation SAC or the Secure HE-LTF Counter; the next instance uses a deliberately invalid Null-SAC HE-LTF and carries the next Validation SAC and counter value in the protected R2I LMR; normal operation then resumes using these updated values.

The standard explicitly states that a station shall discard the Validation SAC used in a failed exchange and shall not use the same SAC again. Since the SAC and the secure HE-LTF keys are derived from the Secure HE-LTF Counter and the seed, this implies a strictly monotonic counter and no reuse of the associated keying material. Under normal conditions, this is straightforward to implement, but error handling, state resets, or firmware bugs can cause implementations to deviate, for example by failing to increment the counter after a recovery round or certain error conditions.

If, contrary to the standard, a device reuses the same counter value for multiple secure HE-LTF instances, it also reuses the same Validation SAC and the same keying material. An attacker who learns or infers these values can then mount the physical-layer attacks on ToA estimation described in Sect.~\ref{sec:distance_attack}, undermining one of the core security goals of secure HE-LTFs. This makes correct implementation of error recovery and strict monotonicity of the Secure HE-LTF Counter a critical requirement for NGP deployments. Similar implementation flaws in counter and nonce handling have already led to practical attacks in related protocols, such as key reinstallation in WPA2 (KRACK) and nonce reuse in TLS with AES-GCM~\cite{vanhoef2017key,bock2016nonce}. While hardware limitations (Sect.~\ref{sec:support}) prevented us from testing this specific recovery mechanism in practice, this history underlines the importance of getting these details right.

\subsection{Denial-of-service considerations}\label{sec:dos}

The configuration choices above not only introduce downgrade risks but also affect availability. If an RSTA requires both a PTK and secure HE-LTFs, an attacker can no longer silently downgrade a session. However, ranging then becomes easy to block entirely: disrupting the initial FTMR or PASN exchange prevents a secure session from being established, and a correctly configured ISTA will refuse to fall back to legacy FTM.

Even once a secure FTM session is active, the error-recovery mechanisms around the Secure HE-LTF Counter and Validation SAC offer further opportunities for disruption. Any manipulation that causes secure HE-LTF validation to fail forces a recovery round in which measurements are deliberately treated as invalid before resuming with updated counter values. By repeatedly triggering such failures in the attacker models of Sect.~\ref{sec:personal-mode} and Sect.~\ref{sec:pasn_sec}, an adversary can degrade or effectively prevent useful measurements without ever completing a single successful ranging exchange.

A strictly configured NGP deployment that correctly refuses insecure fallbacks can be robust against downgrades yet vulnerable to simple availability attacks that require only selective interference with a few protocol messages.
Table~\ref{tab:logical-summary} summarises the three main classes of logical-layer issues.

\begin{table}[b]
  \centering
  \caption{Summary of logical-layer issue classes.}
  \label{tab:logical-summary}
  \scalebox{0.9}{
  \begin{tabular}{p{0.31\linewidth}@{\hspace{0.05\linewidth}}p{0.56\linewidth}}
    \hline
    \textbf{Issue class} & \textbf{Description} \\
    \hline
    Deployment modes
    (Personal, standalone PASN)
    &
    Secure FTM cannot be firmly bound to a uniquely trusted AP; anyone with the shared credentials can become the ranging peer. \\
    \hline
    Configuration and negotiation
    &
    Permissive settings allow downgrades from secure to legacy FTM or to deterministic HE-LTFs without breaking cryptography. \\
    \hline
    Error handling and availability
    &
    Subtle counter/SAC handling and strict policies make secure FTM fragile and easy to disable via selective interference. \\
    \hline
  \end{tabular}}
\end{table}


\section{Physical-Layer Analysis}\label{sec:physical}

In this section, we analyse the physical-layer characteristics of secure FTM in NGP, using the simulation and measurement setup from Sect.~\ref{sec:scope}.
We first study the secure HE-LTF waveform's predictability and how much an attacker can learn from partially observed symbols.
We then evaluate practical distance-reduction attacks under both deterministic and partial-observation adversary models.
Finally, we examine how the zero-power guard interval interacts with realistic RF front ends and spectral-mask constraints.

\subsection{Physical-Layer Signal}
To ensure that our simulation setup is representative, we compared the simulated waveforms against development-board captures and confirmed agreement in the time-domain structure and overall power spectral density.
Legacy FTM in IEEE~802.11mc relies on predictable HE-LTF sequences for ToA estimation.
IEEE~802.11az introduces several changes: the HE-LTF content is randomised, the conventional guard interval (GI) is replaced with a zero-power GI, and a pseudorandom phase rotation is applied per stream.

In the secure format, the HE-LTF sequence is constructed from pseudorandom 64-QAM symbols and per-stream phase rotations on the active subcarriers. Depending on the bandwidth, there are 122, 242, or 498 non-zero subcarriers for 20, 40, and 80~MHz, respectively (higher bandwidths are built from the 80~MHz definition). For 20~MHz, 122 out of 256 subcarriers carry non-zero pseudorandom symbols, with the remaining subcarriers used for lower and upper guard bands, the DC subcarrier, and an alternating null pattern. To avoid repetition in the time domain, the standard uses the 2x HE-LTF type, where only half of the IFFT output is transmitted for each symbol.

\subsection{HE-LTF estimation}
In the time-domain signal, an attacker aims to advance the HE-LTF so the receiver records an earlier ToA.
To mount early-detect/late-commit attack~\cite{multicarrier_security}, in which the attacker observes the start of a symbol and immediately transmits the rest, the attacker must predict the transmitted symbol from a partial observation.
This amounts to solving an underdetermined system of equations.

For an $N$ non-zero subcarrier 64-QAM symbol with pseudorandom symbols, the entropy is $H = N \cdot \log_2 64 = 6 \cdot N$, making a brute-force search impractical even when only part of the symbol is unknown.

A more realistic approach is to infer \emph{soft} information about the transmitted symbols from the partial observation using Bayesian inference, in particular message-passing methods \cite{wymeersch2007iterative,dauwels2007variational}. These algorithms can incorporate structural knowledge such as the constellation diagram and channel statistics.

In a MATLAB simulation, we emulate the secure HE-LTF as an OFDM symbol with $N = 122$ non-zero subcarriers, each populated with a randomly selected 64-QAM symbol $X_k$, $k = 1,\ldots,N$. In addition, a common phase shift $\beta \in \{0,\pi/4\}$ is drawn at random. The corresponding time-domain signal is $\mathbf{y} = \mathbf{F}^H e^{\mathrm{i}\beta} \mathbf{X} + \mathbf{w}$, where $\mathbf{F}$ is the unitary DFT matrix (implemented via an IFFT), $\mathbf{X}$ collects all $X_k$, and $\mathbf{w}$ is white Gaussian noise (AWGN channel). From $\mathbf{y}$, the attacker observes only $M < N$ samples ($\mathbf{y}_\text{obs}$), and uses these to estimate the symbols and predict the unobserved part of the sequence.

The attacker seeks estimates $\hat{\mathbf{X}}$ from $\mathbf{y}_\text{obs}$ in order to reconstruct the full time-domain signal $\hat{\mathbf{y}} = \mathbf{F}^H \hat{\mathbf{X}}$. Following a maximum a posteriori (MAP) criterion, the optimal symbol estimate is
\begin{align}
    \hat{X}_k & = \arg\max_{X_k} p\bigl(X_k \mid \mathbf{y}_\text{obs}\bigr) \,,
    \label{eq:est_Xk}
\end{align}
where the marginal posterior is $p\bigl(X_k \mid \mathbf{y}_\text{obs}\bigr) = \sum_{\beta} \sum_{X_{\sim k}} p\bigl(\mathbf{X}, \beta \mid \mathbf{y}_\text{obs}\bigr)$. Here, $X_{\sim k}$ denotes all $X_j$ with $j \neq k$. By Bayes' rule, $p\bigl(X_k \mid \mathbf{y}_\text{obs}\bigr) \propto \sum_{\beta} \sum_{X_{\sim k}} p\bigl(\mathbf{y}_\text{obs} \mid \mathbf{X}, \beta\bigr) \, p(\mathbf{X}) \, p(\beta)$. Similarly, the phase is estimated as
\begin{align}
    \hat{\beta} & = \arg\max_{\beta} p\bigl(\beta \mid \mathbf{y}_\text{obs}\bigr)\,,
    \label{eq:est_beta}
\end{align}
with $p\bigl(\beta \mid \mathbf{y}_\text{obs}\bigr) \propto \sum_{\mathbf{X}} p\bigl(\mathbf{y}_\text{obs} \mid \mathbf{X}, \beta\bigr) \, p(\mathbf{X}) \, p(\beta)$. Direct marginalisation in \eqref{eq:est_Xk} and \eqref{eq:est_beta} is computationally expensive, requiring on the order of $2^{6(N-1)}$ summations per $\hat{X}_k$. We therefore resort to message passing on a factor graph \cite{wymeersch2007iterative} with Monte Carlo sampling \cite{dauwels2006particle,etzlinger2011message}. For the results below, we run a particle-based message-passing algorithm with 300 samples and set the noise variance such that the SNR is $7\,$dB.

\begin{figure}[!t]
    \centering
    \includegraphics[width=1.01\linewidth]{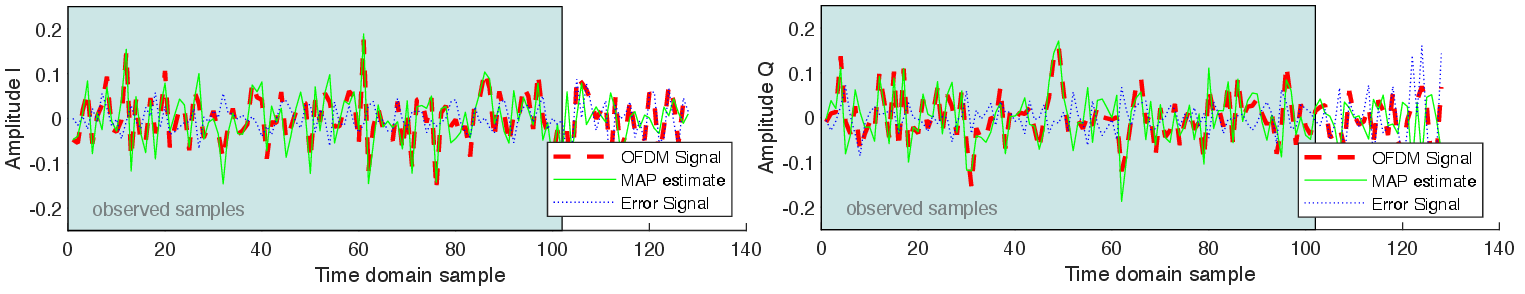}
    \caption{Waveform reconstruction after partial observation (80\%).}
    \vspace{-4mm}
    \label{fig:pred_timedomain_symb}
\end{figure}

After inference, we obtain approximate marginals for each constellation symbol and the common phase, $p\bigl(X_k \mid \mathbf{y}_\text{obs}\bigr)$ and $p\bigl(\beta \mid \mathbf{y}_\text{obs}\bigr)$. Concretely, for each $X_k$ and for $\beta$ we obtain normalised probability masses over the 64-QAM alphabet and the two phase candidates. The MAP estimates in \eqref{eq:est_Xk} and \eqref{eq:est_beta} are given by the maximisers of these distributions. Figure~\ref{fig:pred_timedomain_symb} shows a time-domain symbol where $80\%$ of the samples are observed and used for reconstruction. The reconstructed signal (green) follows the observed samples (red) closely, but the error trace (blue dotted) reveals non-negligible error outside the observation window.

To quantify both accuracy and confidence of the MAP estimate, we report for the symbols: (i) the empirical negative log-likelihood (NLL), defined as $\text{NLL} = -\log \hat{p}(x^\ast)$, where $x^\ast$ is the ground-truth symbol and $\hat{p}(x)$ is the approximated posterior mass; and (ii) the posterior entropy $H = -\sum_x \hat{p}(x) \log \hat{p}(x)$, which decreases as the posterior concentrates around a small set of candidates.

\begin{figure}[!b]
  \centering
  \subfloat[\label{fig:pred_results_a}]{%
    \begin{minipage}[b]{0.48\linewidth}
      \centering
      \includegraphics[width=\linewidth]{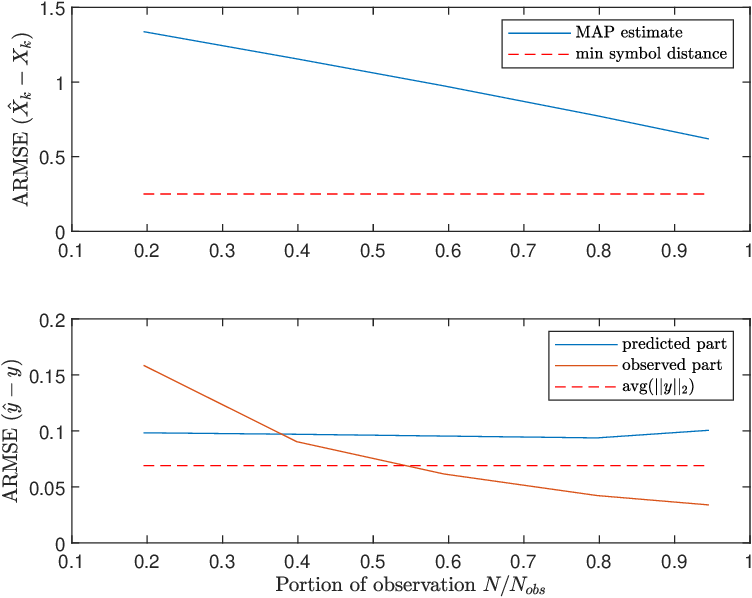}
    \end{minipage}%
  }\hfill
  \subfloat[\label{fig:pred_results_b}]{%
    \begin{minipage}[b]{0.48\linewidth}
      \centering
      \includegraphics[width=\linewidth]{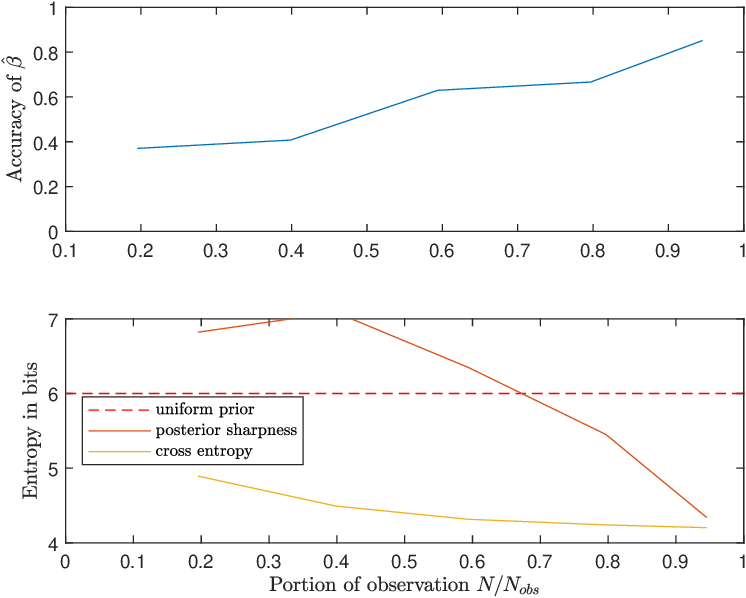}
    \end{minipage}%
  }
  \caption{Simulation results, 25 iterations. (a) Average RMSE. Top: distance between estimated and true constellation symbols, compared to minimum symbol distance. Bottom: average RMSE in time domain for the observed signal part and the predicted signal part, compared to the mean symbol energy of the HE-LTF $y$. (b) Top: accuracy of random phase shift estimation. Bottom: sharpness and cross entropy of the posterior function on the constellation symbols $X_k$.}
  \label{fig:pred_results}
\end{figure}

Figure~\ref{fig:pred_results} summarises the prediction limits under partial observation. In (a), the symbol-domain root mean square error (RMSE) (top) exceeds the minimum constellation distance, indicating that exact recovery of the pseudorandom sequence $\{X_k\}$ from partial samples is infeasible. Furthermore, while time-domain reconstruction (bottom) improves on the observed segment as the window increases, the error on the unobserved continuation remains largely unchanged. Figure~\ref{fig:pred_results_b} highlights that the common phase $\beta$ can be estimated accurately with more observations. Although MAP symbol decisions remain unreliable, the soft posteriors over $X_k$ sharpen with increasing observation (lower entropy and NLL).
This soft information can support short-horizon forecasting or detection, but it does not enable reliable full-sequence prediction beyond the observed window.

\subsection{Distance reduction attack}\label{sec:distance_attack}
Secure HE-LTF repetitions for a given user must be unique, derived from fresh keying material and pseudorandom bits. This symbol freshness is crucial: if errors or permissive configurations (e.g., known keys, counter reuse, or disabled secure HE-LTFs in Sect.~\ref{sec:logical}) make the sequence deterministic, the attacker gains full knowledge of the HE-LTF. This reduces the hard problem of predicting unseen samples to simply generating or replaying exact time-shifted replicas. Distance-reduction attacks then become considerably more practical, especially for correlation-based ToA estimators that readily lock on to the advanced replica rather than treating it as an additional multipath component.



\begin{figure}[!b]
    \centering
    \subfloat[\label{fig:leuven_attack_a}]{%
        \begin{minipage}[b]{0.45\linewidth}
            \centering
            \includegraphics[width=\linewidth]{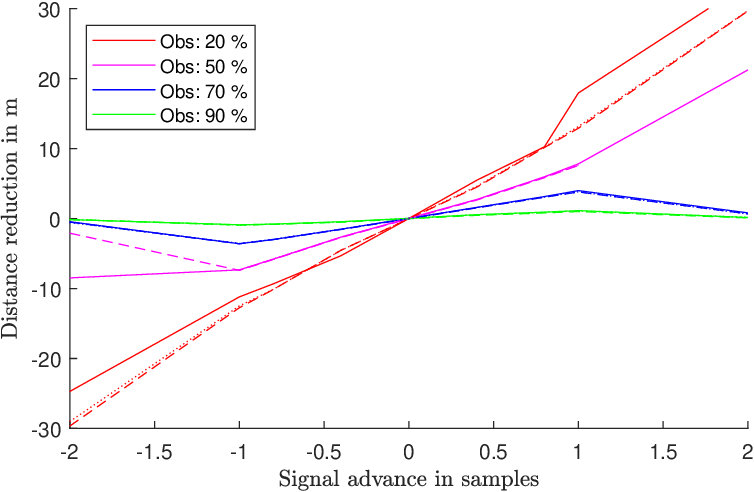}
        \end{minipage}%
    }
    \hspace{3mm}
    \subfloat[\label{fig:leuven_attack_b}]{%
        \begin{minipage}[b]{0.45\linewidth}
            \centering
            \includegraphics[width=\linewidth]{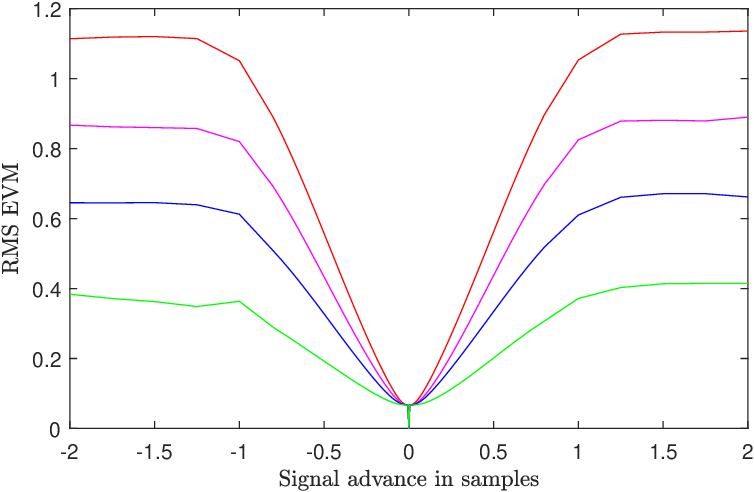}
        \end{minipage}%
    }
    \caption{(a) Distance bias induced by advancing a fraction of the secure HE-LTF. The attacker observes the first $k$\% of the secure HE-LTF to predict and transmit an advanced replica for the remaining portion. The positive x-axis denotes an advance (distance reduction), while negative values denote a delay (distance enlargement). (b) RMS error vector magnitude (EVM) after HE-LTF demodulation versus the applied signal advance.}
    \label{fig:leuven_attack}
    \vspace{-4mm}
\end{figure}

We model an attacker that transmits an advanced waveform segment to bias the HE-LTF ToA estimate. If fully known, for instance by exploiting the vulnerabilities identified in Sect.~\ref{sec:logical}, the attacker can trivially synthesise and inject the entire replica. Here, we evaluate a weaker attacker who must first observe $k$\% of the secure HE-LTF to synthesise an advanced replica for the remainder, cancelling the legitimate component to replace it with the advanced signal. Both scenarios yield a controllable advance of the first path and a proportional distance reduction. We evaluate this bias in MATLAB using a MUSIC-based ToF estimator~\cite{jiokeng2020ftm}. Figure~\ref{fig:leuven_attack_a} plots the distance error versus signal advance for different observation fractions $k$ (e.g., 20\% observed implies 80\% advanced). Solid, dashed, and dotted curves represent SNRs of 5, 15, and 25~dB, respectively.
Delaying (negative advance values) enlarges the distance, while advancing reduces it; larger advanced portions yield larger biases for the same advance.
Crucially, accurately predicting a short segment already induces substantial bias, demonstrating that the weaker attacker succeeds without needing the entire sequence.

A practical limitation is that large advances increasingly distort the OFDM symbol, potentially breaking demodulation. If the sequence is already known, this symbol distortion does not occur. Under partial observation, the post-demodulation RMS EVM rises with the applied advance (Fig.~\ref{fig:leuven_attack_b}). Since reliable 64-QAM operation requires low error vector magnitude (EVM), extreme advances become self-defeating as demodulation fails, or become detectable through degraded PHY metrics. This aligns with IEEE~802.11az's observation that secure HE-LTF attacks introduce interference monitorable via SIR drops across repetitions~\cite[Annex AE]{standard_80211_2024}; however, practical algorithms to reliably detect such drops remain sparse.

\subsection{Zero-power GI}\label{sec:zero-power-gi}

In this section, we examine how the zero-power guard interval interacts with the RF front end and the HE transmit spectral mask requirements~\cite{standard_80211_2024}, focusing on a 20~MHz configuration commonly used for indoor positioning.


\begin{figure}[!b]
  \centering
  \subfloat[Legacy HE-LTF with cyclic prefix.\label{fig:zero-power-a}]{%
    \begin{minipage}[b]{0.49\textwidth}
      \centering
      \includegraphics[scale=0.39]{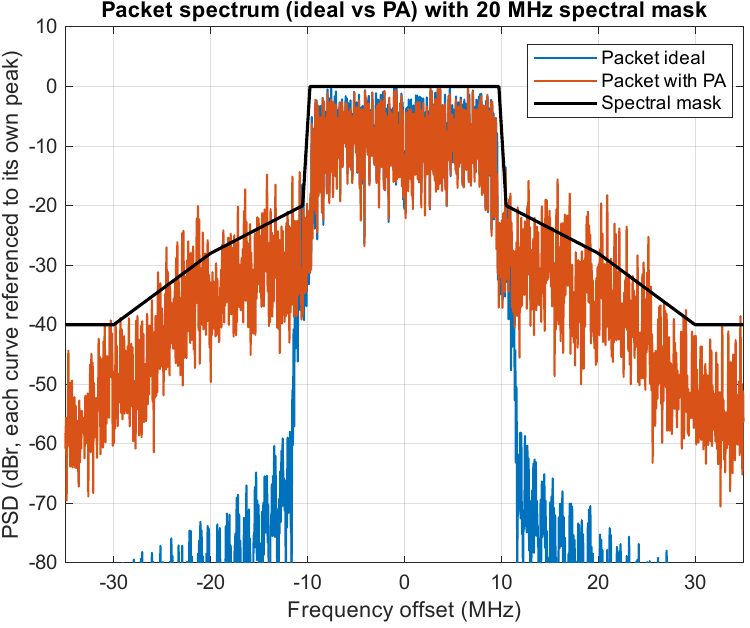}
    \end{minipage}%
  }\hfill
  \subfloat[Secure HE-LTF with zero-power GI.\label{fig:zero-power-b}]{%
    \begin{minipage}[b]{0.49\textwidth}
      \centering
      \includegraphics[scale=0.39]{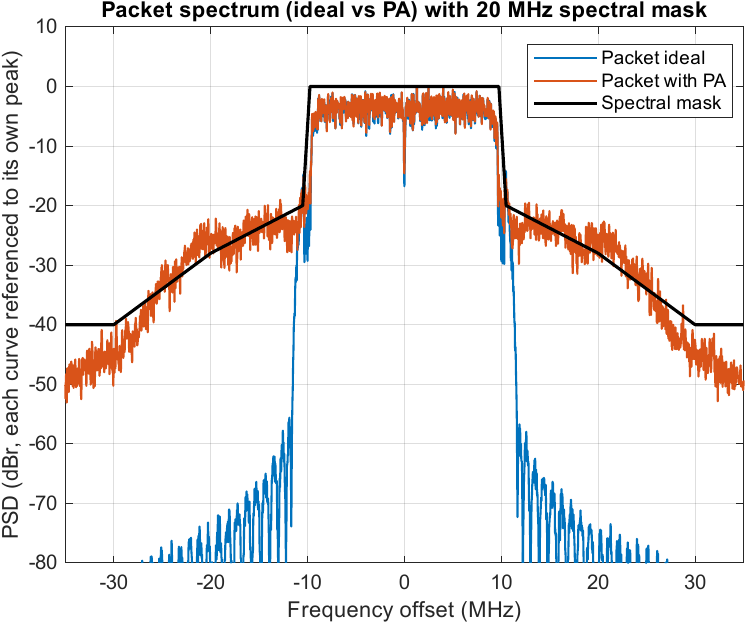}
    \end{minipage}%
  }
  \caption{Simulated power spectral density of 20~MHz NDPs for the non-secure (left) and secure (right) variants, before (blue) and after (orange) the power amplifier model, compared against the standard-defined spectral mask (black).}
  \label{fig:zero-power}
  \vspace{-4mm}
\end{figure}

Starting from the IEEE~802.11az waveform generation example~\cite{mathworks80211azGen}, we generate ranging NDP waveforms for two cases: (i) a legacy HE-LTF format with a cyclic prefix and (ii) a secure HE-LTF format with a zero-power guard interval. To approximate a practical transmitter, we pass each signal through a memoryless Rapp power-amplifier model~\cite{Rapp} with smoothness parameter $p = 4$ and saturation level $A_\text{sat} = 0.7$, chosen to match the output spectrum of a legacy 20~MHz ranging waveform measured on our boards. From the resulting output signals, we compute the power spectral density (PSD) and compare it to the 20~MHz interim transmit spectral mask defined in the standard~\cite{standard_80211_2024}.

Figure~\ref{fig:zero-power} shows the resulting PSDs. The blue curves (ideal baseband) stay within the 20~MHz spectral mask, as expected. The orange curves (after the Rapp amplifier) exhibit spectral regrowth in both cases.

To quantify the impact of the zero-power GI, we evaluate the margin to the mask over the $-30$ to $30$~MHz offset range. For the non-secure variant, the worst-case excess over the mask is about $+10.9$~dB, the 99th percentile margin about $+5.9$~dB, and roughly $6.7\%$ of frequency bins exceed the mask. For the secure variant, the worst-case excess drops to about $+6.2$~dB and the 99th percentile margin to about $+4.8$~dB, but the fraction of bins above the mask rises to approximately $31.6\%$. In other words, the zero-power guard interval tends to spread out-of-band energy more evenly towards the mask: individual peaks are reduced, but a much larger portion of the spectrum operates close to, or above, the interim spectral limit.





\section{Discussion and Recommendations}\label{sec:discussion}
In this section, we reflect on the combined findings from our analyses and discuss their implications for secure FTM in NGP. We then relate these to the limited deployment of secure FTM and to IEEE~802.11bk.

\subsection{Security implications}\label{sec:sec_implications}

NGP is intended to support applications ranging from in-store navigation and analytics to keyless entry and proximity-based access control~\cite{IEEE80211az}. Our logical-layer analysis indicates that, in configurations that are likely to dominate in practice, such as WPA2/WPA3-Personal networks and standalone PASN deployments (Sect.~\ref{sec:logical}), secure FTM cannot reliably attest proximity to a uniquely trusted entity. Any party who knows the passphrase or can complete the PASN exchange can act as the apparent ranging peer. This makes high-stakes use cases, such as physical access control or distance-based authorisation, hard to justify when they rely directly on secure FTM in these environments.

Secure FTM can still be useful in such settings for low-stakes tasks, for example, indoor navigation, coarse localisation, or non-safety-critical analytics, provided it is treated as a best-effort signal rather than a strict security boundary. In contrast, deployments that require strong guarantees about \emph{which} device is being ranged must anchor secure FTM in robust mutual authentication, for example via 802.1X-based Enterprise networks, and must enforce strict policies that require both a PTK and secure HE-LTFs and disallow fallback to legacy modes. As Sect.~\ref{sec:dos} shows, this hardening improves integrity but also makes secure FTM more susceptible to simple denial-of-service.

The physical-layer results sharpen these conclusions. When secure HE-LTFs are implemented as specified, with fresh keying material, monotonic counters, and pseudorandom symbols, partial observations yield only soft information. If secure HE-LTFs are disabled, counters are reused, or the same training symbols reappear across instances (Sect.~\ref{sec:error-recovery}), the problem collapses to recording one symbol and replaying time-shifted replicas, enabling correlation-based distance reduction. Overall, the intended security of NGP only materialises when authentication, configuration, and physical-layer behaviour all satisfy relatively strong assumptions; otherwise, the effective guarantees are significantly weaker than the design suggests. The same observations apply to IEEE~802.11bk, which reuses NGP’s security mechanisms.

\subsection{Limited support and deployment challenges}\label{sec:dis_support}

In Sect.~\ref{sec:support}, we observed that secure FTM support in commercial devices and development platforms is scarce. Our analysis suggests technical and economic reasons for this slow adoption.

\textbf{Physical-layer constraints and spectral masks.}
With realistic power-amplifier nonlinearities, the secure HE-LTF with a zero-power GI operates closer to, and in parts above, the transmit spectral mask than the legacy cyclic-prefix waveform (Sect.~\ref{sec:zero-power-gi}). Staying within both the standard mask and regulatory limits may require additional linearity margin, lower transmit power, or stronger filtering, which increases RF complexity compared to legacy FTM.

\textbf{Zero-power GI implementation cost.}
A true zero-power GI requires fast, well-controlled switching of the transmit chain so that the output power drops and recovers within a single OFDM symbol without excessive spectral splatter or timing distortion. Analogue front ends that were designed around a conventional cyclic prefix are unlikely to meet these constraints with firmware changes alone, so full zero-power GI support tends to require non-trivial hardware adaptations.

\textbf{Baseband life cycles and business incentives.}
If secure HE-LTFs and zero-power GIs cannot be added purely in software, vendors need new or updated baseband silicon, with multi-year design and qualification cycles. Our logical-layer results indicate that the strongest security benefits arise only in tightly managed Enterprise-style deployments, while common Personal-mode networks cannot safely support the most ambitious use cases. This combination of hardware cost and comparatively narrow high-assurance use cases reduces the incentive to prioritise secure FTM over existing legacy FTM support.

Taken together, these factors suggest that the limited availability of secure FTM is not merely a delay in adoption, but reflects genuine RF, hardware, and deployment hurdles that must be addressed before NGP can deliver its intended security properties at scale.

\subsection{Recommendations}\label{sec:recommendations}

In its current form, secure FTM is better suited to low- and medium-stakes applications than to high-assurance access control.
In common WPA2/WPA3-Personal and standalone PASN deployments (Sect.~\ref{sec:logical}), secure FTM cannot reliably attest proximity to a uniquely trusted entity.
We therefore advise against using it as the primary security anchor for keyless entry or distance-based authorisation.
Instead, secure FTM is appropriate for indoor navigation or non-safety-critical analytics, where ranging is treated as a best-effort signal rather than a strict security boundary.

Where stronger guarantees are required, secure FTM must be rooted in mutual authentication (e.g., 802.1X-based Enterprise networks) alongside strict configuration policies.
Devices supporting secure FTM should default to requiring a PTK, avoid EDCA-based legacy fallbacks when trigger-based modes are available, and honour secure HE-LTF requirements.
Furthermore, error recovery mechanisms must maintain strictly monotonic counters and prevent keying material reuse (Sect.~\ref{sec:error-recovery}) to avoid undermining physical-layer protections.

For standardisation and certification, we recommend explicitly mapping deployment models to supported use cases, distinguishing environments that tolerate Personal mode from those requiring Enterprise-grade mutual authentication.
Finally, given the demanding RF requirements of the zero-power GI (Sect.~\ref{sec:zero-power-gi}), future amendments should explore hybrid designs such as a reduced-power pseudorandom sequence.
This avoids both deterministic cyclic prefixes and harsh RF switching, easing spectral-mask compliance and lowering barriers to adoption.


\section{Related Work}\label{sec:related-work}

\subsubsection{Security of Wi-Fi FTM and multicarrier ranging.}
Early security analyses focus on the original, non-secure IEEE~802.11mc FTM procedures. Schepers et al.\ provide an end-to-end analysis demonstrating practical distance manipulation and spoofing attacks, and discuss protocol-level mitigations~\cite{wifi_80211mc_security1}, while Henry et al.\ show how attackers bias location estimates in GPS-like settings~\cite{wifi_80211mc_security2}. Subsequent work benchmarks FTM accuracy on IoT devices~\cite{wifi_80211mc_benchmarking}, explores privacy risks and parameter exposures~\cite{wifi_positioning_privacy}, and proposes crowd-wisdom defences for 802.11mc and early 802.11az deployments~\cite{wifi_80211az_crowd}. Complementing this, physical-layer studies demonstrate how deterministic OFDM training and cyclic prefixes enable precise distance-reduction and extension attacks~\cite{multicarrier_security}, prompting the use of super-resolution estimators like frequency-domain MUSIC~\cite{toa_estimation_music}. While these works highlight critical vulnerabilities and multicarrier exploitation strategies, they largely predate the finalised NGP mechanisms and treat the physical layer as an abstract ranging primitive.

\subsubsection{Positioning performance and deployments of secure FTM.}
With the standardisation of NGP, recent literature has shifted towards positioning performance. Works evaluate 802.11az-based ranging at millimetre-wave frequencies~\cite{wifi_positioning_mmwave}, explore parameter choices and communication capabilities~\cite{wifi_80211az_capabilities}, and survey indoor positioning across the 802.11mc/az/bk amendments~\cite{wifi_positioning_survey}. Concurrently, industry documentation illustrates growing FTM adoption: Android exposes Wi-Fi RTT support via its APIs~\cite{wifi_rtt}, Google documents 802.11mc-compatible devices~\cite{android_wifi_rtt}, and providers like IndoorAtlas advertise FTM-backed solutions~\cite{indooratlas}. However, these academic and industrial efforts focus primarily on legacy FTM or performance metrics, offering limited analysis of protocol robustness against active adversaries. To the best of our knowledge, no peer-reviewed work provides a combined protocol- and waveform-level security analysis of the finalised secure FTM procedures in IEEE~802.11az/bk, nor quantifies the extent to which these features are exposed in currently deployed commercial and development hardware.

 \section{Conclusion}\label{sec:conclusion}

In this paper, we examined the security and adoption of secure Wi-Fi ranging as defined in IEEE~802.11az, combining standards analysis with simulations and measurements on commercial and development hardware.

At the logical layer, we showed that common WPA2/WPA3-Personal and standalone PASN deployments cannot reliably attest proximity to a uniquely trusted entity, remaining vulnerable to unauthenticated ranging, downgrades, and denial-of-service. At the physical layer, we analysed secure waveform predictability, the impact of symbol repetition, and how the zero-power guard interval interacts with spectral masks under realistic RF nonlinearity.

Our results indicate that secure Wi-Fi ranging is highly sensitive to configuration choices and difficult to integrate under current RF and deployment constraints, which helps to explain its limited support in existing devices. We outlined where secure FTM can reasonably be used today and provided recommendations for vendors and standardisation bodies to clarify security guarantees and reduce implementation barriers in future profiles and revisions.

\vspace{-3.5mm}

\begin{credits}
\subsubsection{\ackname}
This work was supported in part by the Flemish Government through the Cybersecurity Research Program (VOEWICS02), and in part by COST Action CA22168 – Physical Layer Security for Trustworthy and Resilient 6G Systems (6G-PHYSEC).

\end{credits}
%
%
\bibliographystyle{splncs04}

\bibliography{biblio}

@inproceedings{etzlinger2011message,
  title={{Message passing methods for factor graph based MIMO detection}},
  author={Etzlinger, Bernhard and Haselmayr, Werner and Springer, Andreas},
  booktitle={2011 Wireless Advanced},
  year={2011},
  organization={IEEE}
}

@inproceedings{dauwels2006particle,
  title={Particle methods as message passing},
  author={Dauwels, Justin and Korl, Sascha and Loeliger, Hans-Andrea},
  booktitle={2006 IEEE International Symposium on Information Theory},
  year={2006},
  organization={IEEE}
}

@book{wymeersch2007iterative,
  title={{Iterative Receiver Design}},
  author={Wymeersch, Henk},
  publisher={Cambridge University Press},
  address={Cambridge},
  year={2007}
}

@inproceedings{dauwels2007variational,
  title={On variational message passing on factor graphs},
  author={Dauwels, Justin},
  booktitle={2007 IEEE international symposium on information theory},
  pages={2546--2550},
  year={2007},
  organization={IEEE}
}

@inproceedings{wifi_80211mc_security1,
  title={{Here, there, and everywhere: Security analysis of Wi-Fi Fine Timing Measurement}},
  author={Schepers, Domien and Singh, Mridula and Ranganathan, Aanjhan},
  booktitle={Proceedings of the 14th ACM Conference on Security and Privacy in Wireless and Mobile Networks},
  year={2021}
}

@inproceedings{wifi_80211mc_security2,
  title={{Ranging and Location attacks on 802.11 FTM}},
  author={Henry, Jerome and Busnel, Yann and Ludinard, Romaric and Montavont, Nicolas},
  booktitle={2021 IEEE 32nd Annual International Symposium on Personal, Indoor and Mobile Radio Communications (PIMRC)},
  pages={1481--1486},
  year={2021},
  organization={IEEE}
}

@inproceedings{multicarrier_security,
  title={{Security of multicarrier time-of-flight ranging}},
  author={Leu, Patrick and Kotuliak, Martin and Roeschlin, Marc and Capkun, Srdjan},
  booktitle={Proceedings of the 37th Annual Computer Security Applications Conference},
  pages={887--899},
  year={2021}
}

@article{wifi_positioning_mmwave,
  title={{IEEE 802.11az Indoor Positioning with mmWave}},
  author={Picazo-Mart{\'\i}nez, Pablo and Barroso-Fern{\'a}ndez, Carlos and Mart{\'\i}n-P{\'e}rez, Jorge and Groshev, Milan and de la Oliva, Antonio},
  journal={IEEE Communications Magazine},
  volume={62},
  number={10},
  pages={126--131},
  year={2023},
  publisher={IEEE}
}

@inproceedings{wifi_80211mc_benchmarking,
  title={{Benchmarking and Security Considerations of Wi-Fi FTM for Ranging in IoT Devices}},
  author={Singh, Govind and Pandey, Anshul and Prakash, Monika and Andreoni, Martin and Baddeley, Michael},
  booktitle={Proceedings of Cyber-Physical Systems and Internet of Things Week 2023},
  year={2023}
}

@article{wifi_positioning_survey,
  title={{Indoor Positioning with Wi-Fi Location: A Survey of IEEE 802.11 mc/az/bk Fine Timing Measurement Research}},
  author={Kosek-Szott, Katarzyna and Szott, Szymon and Ciezobka, Wojciech and Wojnar, Maksymilian and Rusek, Krzysztof and Segev, Jonathan},
  journal={Computer Communications},
  year={2025}
}

@article{wifi_positioning_privacy,
  title={{Privacy-preserving positioning in Wi-Fi Fine Timing Measurement}},
  author={Schepers, Domien and Ranganathan, Aanjhan},
  journal={Proceedings on Privacy Enhancing Technologies},
  year={2022}
}

@inproceedings{wifi_80211az_capabilities,
  title={{Unlocking the Potential of IEEE 802.11az: A Deep Dive into Ranging Capabilities}},
  author={Famili, Alireza and Atalay, Tolga and Stavrou, Angelos},
  booktitle={2025 International Conference on Computing, Networking and Communications (ICNC)},
  pages={763--769},
  year={2025},
  organization={IEEE}
}

@inproceedings{wifi_80211az_crowd,
  title={{Reducing FTM ranging and location attack exposure with crowd-wisdom}},
  author={Henry, Jerome and Busnel, Yann and Ludinard, Romaric and Montavont, Nicolas},
  booktitle={IPIN 2021: 9th International Conference on Indoor Positioning and Indoor Navigation},
  pages={1--16},
  year={2021}
}

@article{toa_estimation_music,
  title={{Super-resolution TOA estimation with diversity for indoor geolocation}},
  author={Li, Xinrong and Pahlavan, Kaveh},
  journal={IEEE transactions on wireless communications},
  volume={3},
  number={1},
  pages={224--234},
  year={2004},
  publisher={IEEE}
}

@ARTICLE{standard_80211_2024,
  author = {{IEEE Standards Association et al.}},
  title  = {{IEEE Standard for Local and Metropolitan Area Networks--Part 11: Wireless LAN Medium Access Control (MAC) and Physical Layer (PHY) Specifications}},
  journal={IEEE Std 802.11-2024}, 
  year   = {2025}
}

@ARTICLE{standard_80211bk,
  author = {{IEEE Standards Association et al.}},
  title  = {{IEEE Standard for Local and Metropolitan Area Networks--Part 11: Wireless LAN MAC and PHY Specifications Amendment 3: 320MHz Positioning}},
  journal={IEEE Std 802.11bk-2025},
  year   = {2025}
}

@misc{android_wifi_rtt,
  title        = {{Wi-Fi location: Ranging with RTT}},
  author       = {{Android}},
  howpublished = {\url{https://developer.android.com/develop/connectivity/wifi/wifi-rtt#supported-devices}},
  note         = {Accessed 7 January 2026}
}

@misc{wifirttlocator,
  title        = {{WifiRttLocator App}},
  author       = {{Google LLC}},
  organization = {Google Play},
  url          = {https://play.google.com/store/apps/details?id=com.google.android.apps.location.rtt.wifirttlocator},
  note         = {Accessed 7 January 2026},
}

@misc{indooratlas,
  title        = {{Unlock Smart Spaces with IndoorAtlas}},
  author       = {{IndoorAtlas}},
  howpublished = {\url{https://www.indooratlas.com/}},
  note         = {Accessed 7 January 2026},
  year         = {2026},
}

@misc{wifi_rtt,
  title        = {{Wi-Fi RTT}},
  author       = {{Android Open Source Project}},
  howpublished = {\url{https://source.android.com/docs/core/connect/wifi-rtt}},
  note         = {Accessed 7 January 2026},
  year         = {2026},
}

@article{dolev2003security,
  title={{On the Security of Public Key Protocols}},
  author={Dolev, Danny and Yao, Andrew},
  journal={IEEE Transactions on Information Theory},
  volume={29},
  number={2},
  pages={198--208},
  year={1983},
  publisher={IEEE}
}

@inproceedings{jiokeng2020ftm,
  title={{When FTM discovered MUSIC: Accurate WiFi-based ranging in the presence of multipath}},
  author={Jiokeng, Kevin and Jakllari, Gentian and Tchana, Alain and Beylot, Andr{\'e}-Luc},
  booktitle={IEEE INFOCOM 2020-IEEE Conference on Computer Communications},
  pages={1857--1866},
  year={2020},
  organization={IEEE}
}

@misc{mathworks80211azToA,
  author       = {{MathWorks}},
  title        = {{802.11az Positioning Using Super-Resolution Time of Arrival Estimation}},
  howpublished = {\url{https://www.mathworks.com/help/wlan/ug/802-11az-indoor-positioning-using-super-resolution-time-of-arrival-estimation.html}},
  note         = {Accessed 13 January 2026}
}

@misc{mathworks80211azGen,
  author       = {{MathWorks}},
  title        = {{802.11az Waveform Generation}},
  howpublished = {\url{https://www.mathworks.com/help/wlan/ug/802-11az-waveform-generation.html}},
  note         = {Accessed 13 January 2026},
  year         = {2026}
}

@misc{IEEE80211az,
  author       = {{IEEE Standards Association}},
  title        = {{Newly Released {IEEE} 802.11az Standard Improving Wi-Fi Location Accuracy is Set to Unleash a New Wave of Innovation}},
  howpublished = {\url{https://standards.ieee.org/beyond-standards/newly-released-ieee-802-11az-standard-improving-wi-fi-location-accuracy-is-set-to-unleash-a-new-wave-of-innovation/}},
  note         = {Accessed 26 January 2026},
  year         = {2026}
}

@inproceedings{Dragonblood2019,
  title={{Dragonblood: Analyzing the Dragonfly Handshake of WPA3 and EAP-pwd}},
  author={Vanhoef, Mathy and Ronen, Eyal},
  booktitle={2020 IEEE Symposium on Security and Privacy (SP)},
  pages={517--533},
  year={2020},
  organization={IEEE}
}

@inproceedings{vanhoef2024_twin,
  title={{A security analysis of WPA3-PK: Implementation and precomputation attacks}},
  author={Vanhoef, Mathy and Robben, Jeroen},
  booktitle={International Conference on Applied Cryptography and Network Security},
  pages={217--240},
  year={2024},
  organization={Springer}
}

@inproceedings{vanhoef2017key,
  title={{Key reinstallation attacks: Forcing nonce reuse in WPA2}},
  author={Vanhoef, Mathy and Piessens, Frank},
  booktitle={Proceedings of the 2017 ACM SIGSAC conference on computer and communications security},
  pages={1313--1328},
  year={2017}
}

@inproceedings{bock2016nonce,
  title={{Nonce-Disrespecting adversaries: Practical forgery attacks on GCM in TLS}},
  author={B{\"o}ck, Hanno and Zauner, Aaron and Devlin, Sean and Somorovsky, Juraj and Jovanovic, Philipp},
  booktitle={10th USENIX Workshop on Offensive Technologies (WOOT 16)},
  year={2016}
}

@INPROCEEDINGS{Rapp,
  author={Jayati, Ari Endang and Sipan, Muhammad},
  booktitle={2020 Third International Conference on Vocational Education and Electrical Engineering (ICVEE)}, 
  title={{Impact of Nonlinear Distortion with the Rapp Model on the GFDM System}}, 
  year={2020}
}

\end{document}